\documentclass[useAMS,usenatbib]{mn2e}
\usepackage{amsmath,fleqn,graphicx,amssymb,xcolor}
\usepackage{multirow}
\renewcommand{\[}{\begin{equation}}
\renewcommand{\]}{\end{equation}}
\def\p{\partial}\def\i{{\rm i}}

\def\rd{}

\let\boldgrk=\gkvecten
\let\boldgrksc=\gkvecseven

\def\gkthing#1{{\mathchoice%
	{\hbox{{\boldgrk\char#1}}}
	{\hbox{{\boldgrk\char#1}}}
	{\hbox{{\boldgrksc\char#1}}}
	{\hbox{{\boldgrksc\char#1}}}}}

\def\vtheta{\gkthing{18}}

{\newif\ifnotend
\notendtrue
\def\veclist{ABCDEFGHIJKLMNOPQRSTUVWXYZabcdefghijklmnopqrstuvwxyz.}
\def\top#1#2.{#1}
\def\tail#1#2.{#2.}
\loop\expandafter\xdef\csname v\expandafter\top\veclist\endcsname%
{{\noexpand\bf\expandafter\top\veclist}}
\edef\veclist{\expandafter\tail\veclist}
\if\veclist.\notendfalse\fi\ifnotend\repeat}
{\newif\ifnotend
\notendtrue
\def\callist{ABCDEFGHIJKLMNOPQRSTUVWXYZ.}
\def\top#1#2.{#1}
\def\tail#1#2.{#2.}
\loop\expandafter\xdef\csname c\expandafter\top\callist\endcsname%
{{\noexpand\cal\expandafter\top\callist}}
\edef\callist{\expandafter\tail\callist}
\if\callist.\notendfalse\fi\ifnotend\repeat}
\def\rhop{\rho^{(\alpha)}}
\def\Phip{\Phi^{(\alpha)}}

\def\rhopp{\rho^{(\alpha')}}\def\Phips{\Phi^{(\alpha)*}}

\def\bra#1{\langle#1|}

\def\la{\langle}
\def\ra{\rangle}

\def\d{{\rm d}}

\def\bolOm{\mbox{\boldmath$\Omega$}}
\def\vOmega{\bolOm}

\def\e{\mathrm{e}}\def\s{\mathrm{s}}
\def\vnabla{{\bf\nabla}}

\def\fracj#1#2{{\textstyle{#1\over#2}}}

\def\eqrf#1{(\ref{#1})}

\title[Modes of a stellar system II: non-ergodic systems]
{Modes of a stellar system II: non-ergodic systems}

\author[Jun Yan Lau \& James Binney]{
  Jun Yan Lau$^{1,2}$\thanks{E-mail: jun.lau.20@ucl.ac.uk} \& James Binney$^2$\thanks{E-mail:
  binney@physics.ox.ac.uk}\\   
  $^1$UCL Mullard Space Sciences Laboratory, Holmbury St Mary, Surrey RH5
  6NT\\
  $^2$Rudolf Peierls Centre for Theoretical Physics, Clarendon Laboratory,
  Parks Road, Oxford, OX1 3PU, UK
}

\begin{document}
\maketitle

\begin{abstract}
An equation is derived for the energy of a small disturbance in a system that
is generated by a distribution function (DF) of the form $f(\vJ)$ -- most
galaxies and star clusters can be closely approximated by such a DF. The
theory of van Kampen modes is  extended to such general systems. A {\rd bilinear
form on} the space of DFs is defined such that the energy of a disturbance
is its norm under this form. It is shown that van Kampen modes that
differ in frequency are then orthogonal, with the consequence that the
energies of van Kampen modes are additive. Consequently, most of the insight
into he dynamics of ergodic systems that was gained in a recent paper on the
van Kampen modes of ergodic systems applies to real clusters and galaxies.
\end{abstract}


\section{Introduction}

Stellar systems have hitherto been modelled in mean-field limit of an infinite number
of constituent particles, when
fluctuations vanish. That fluctuations play an essential role in the
evolution of clusters was recognised over half a century ago (Henon, Spitzer,
Chandra), but even now observations are fitted to mean-field models such as
Michie-King models \citep{Michie1963,King1966, McLaughlin2006,Piotto2015,Claydon2019}.

Some fluctuations are internally generated by the shot noise
inherent in a system with a finite number of particles, while other
fluctuations are externally  stimulated by the gravitational fields of
neighbouring systems. As observational data become more precise, there must
come a point at which fluctuations of either type can be detected. Detection
of fluctuations would open the exciting possibility of using galaxies and
star clusters to detect the passage of dark-matter haloes because the tidal
fields of such haloes will excite the large-scale modes of globular clusters
and dwarf spheroidal galaxies.

Modelling the small-scale effects of shot noise is rather straightforward
because the system's self-gravity is only important on the largest scales
\citep[e.g][]{FouvryHRP2021}. But following the derivation of the Balescu-Lenard equation
by \cite{Heyvaerts2010} and \cite{Chavanis2012}, it has become evident that
self-gravity has a big impact on system-scale fluctuations even when the
latter are stimulated by Poisson noise
\citep{FouvryPMC2015,Hamilton2018,LauBinney2019,LauBinney2021err,Heggie2020}. One expects
externally generated fluctuations to be predominantly large-scale, so a
viable theory of fluctuations must encompass self-gravity.

The standard approach to a theory of fluctuations is via normal modes:
By linearising the equations of motion one derives a set of harmonically
oscillating disturbances that is complete in the sense that any initial
condition can be expressed as a linear combination of modes. This
decomposition simultaneously characterises the initial condition in an
physically significant way, and provides a convenient way to compute the
system's evolution by taking advantage  of the almost trivial rule for
evolving a normal mode.

The stability of stellar systems has traditionally been investigated by
determining the frequencies of `Landau modes' because the system is unstable
if any of these frequencies has a positive imaginary part. Landau modes,
however, lack the essential completeness property of normal modes. In the
case of a homogeneous electrostatic plasma, \cite{vanKampen1955} presented the true normal
modes, now known as van Kampen modes in his honour, and \cite{Case1959} proved that van Kampen's modes
are complete. Specification of the distribution function (DF) $f(x,v)$ of a
collisionless system requires much more information than is required to
specify the state of a fluid, and this fact is reflected in key differences
between van Kampen modes and the modes of a fluid system. Crucially, the spectrum of
van Kampen modes contains a continuum of real frequencies, and the DFs of individual modes
contains a Dirac $\delta$-function associated with its frequency. Expression of an
initial condition as a sum of normal modes involves an integral over
frequency that eliminates the $\delta$-function. 

Recently in the first paper in this series \citep[][hereafter Paper
I]{LauBinney2021} we derived the van Kampen modes of ergodic stellar systems,
that is ones with DFs $f(H)$, where $H=\fracj12v^2+\Phi(\vx)$ is the
Hamiltonian of an individual star. They showed that the energy of a
disturbance is the sum of the energies of its constituent van Kampen modes,
and that when the system is stable the energies of all modes are inherently
positive. In this paper we generalise these results to non-ergodic
stellar systems, that is, systems with DF $f(\vJ)$, where $\vJ$ is the vector
of action integrals in the unperturbed system. This class of systems is
extremely broad and includes systems that are flattened or spherical and
exhibit velocity anisotropy and/or systematic rotation. Some our results will
be restricted to the subclass of  systems that are unchanged by reversing all
velocities, which in practice
excludes systems with systematic rotation.

In Section~\ref{sec:math} we introduce the notation and some results that
will be required in later sections. In Section~\ref{sec:E} we extend to
systems with DF $f(\vJ)$ the formula for the energy of a fluctuation that
\cite{NelsonTremaine1999} derived for ergodic systems. In
Section~\ref{sec:mode}
we obtain results for  the van Kampen modes of systems with a DF $f(\vJ)$
that are analogues of results derived for ergodic systems in Paper I. The
route taken to these results does, however, differ significantly to that used
in Paper I, being simpler but less powerful -- the fourth paper in this series
will present the rather intricate generalisation to non-ergodic systems of
the approach used in Paper I, which exploits the Hermitian operator
introduced by \cite{Antonov1961}.  In Section~\ref{sec:discuss} we discuss
the differences between ergodic and non-ergodic systems and the relation
between
true modes and Landau modes.

\section{Mathematical background}\label{sec:math}

Here we introduce essential mathematical tools and establish our notation.
We focus on systems with integrable mean-field potentials,
so their DFs can be written in the form $f(\vJ)$.

\subsection{Angle-action variables} The role that Cartesian variables play for
homogeneous systems is played for spheroidal systems by angle-action variables
$(\vtheta,\vJ)$. The actions $J_i$ are constants of motion while their
conjugate variables, the angles $\theta_i$, increase linearly in time, so
$\vtheta(t)=\vtheta(0)+\vOmega t$. The particles' Hamiltonian $H(\vx,\vv)$ is a
function $H(\vJ)$ of the actions only and the frequencies $\Omega_i$ that
control the rates of increase of the angles are given by $\vOmega=\p
H/\p\vJ$. Angle-action variables are canonical, so the volume element of
phase space $\d^6\vw=\d^3\vx\d^3\vv=\d^3\vtheta\d^3\vJ$ and Poisson brackets can be computed
as
\[
[f,g]=\sum_i\bigg(
{\p f\over\p\theta_i}{\p g\over\p J_i}-{\p f\over\p
J_i}{\p g\over\p\theta_i}\bigg).
\]
Functions on phase space can be expressed as Fourier series:
\[\label{eq:defsFT}
h(\vw)=\sum_\vn h_\vn(\vJ)\e^{\i\vn\cdot\vtheta}\ ;\ 
h_\vn(\vJ)=\int{\d^3\vtheta\over(2\pi)^3}\e^{-\i\vn\cdot\vtheta}h(\vw).
\]
Note that for real $h$, $h_{-\vn}=h_\vn^*$.

\subsection{Potential-density pairs} Following
\cite{Kalnajs1976} we solve Poisson's equation by introducing a basis of
{biorthogonal potential-density} pairs.  That is, a set of pairs
$(\rhop,\Phip)$ such that
\[\label{eq:Poisson}
4\pi G\rhop=\nabla^2\Phip\quad\hbox{and}\quad
\int\d^3\vx\,\Phips\rhopp =-\cE\delta_{\alpha\alpha'},
\]
 where $\cE$ is an arbitrary constant with the dimensions of energy.
Given a density
distribution $\rho(\vx)$, we expand it in the basis
\[\label{eq:defsAp}
\rho(\vx)=\sum_\alpha A_\alpha\rhop(\vx)\quad\Leftrightarrow\quad
\Phi(\vx)=\sum_\alpha A_\alpha\Phip(\vx),
\]
where 
\[\label{eq:Aalpha}
A_\alpha=-{1 \over\cE}\int\d^6\vw\,\Phips(\vx)f(\vw).
\]
 If $\rho$ and $\Phi$ are time-dependent, the $A_\alpha$ become
time-dependent.  From equations \eqrf{eq:Poisson} and \eqrf{eq:defsAp} one
can obtain an expression for $\Phi$ in terms of $\rho$. Comparison of this
relation with Poisson's integral, yields
\[\label{eq:Gxmx}
{G\over|\vx'-\vx|}={1\over\cE}\sum_\alpha\Phip(\vx)\Phips(\vx').
\]
The system's potential energy is
\begin{align} \label{eq:PE}
P&=\fracj12\int\d^3\vx\,\Phi\rho=\fracj12\int\d^3\vx\,\sum_\alpha
A_\alpha\Phi^{(\alpha)}\rho\cr
&=-{\cE\over2}\sum_\alpha|A_\alpha|^2.
\end{align}
A related calculation is the potential energy of one system when placed in
the potential of another. This is
\[
P'=-G\int\d^3\vx\,{\rho(\vx)\rho'(\vx')\over|\vx-\vx'|}.
\]
The symmetry of this expression establishes that the energy of system a in
the potential of system b is the same as that of system b placed in the
potential of system a. Trivial adaptation of the derivation of equation
\eqrf{eq:PE} shows that 
\[\label{eq:PE2}
P'=-\cE\sum_\alpha A^*_\alpha A'_\alpha.
\]
Now
\begin{align}\label{eq:PE3}
P'&=\int\d^6\vw\,f\Phi'=\int\d^3\vJ\,\d^3\vtheta\,\sum_{\vm\vn}
f_\vm\Phi'_\vn\e^{\i(\vm+\vn)\cdot\vtheta}\cr
&=(2\pi)^3\sum_\vn\int\d^3\vJ\, f^*_\vn\Phi'_\vn.
\end{align}

\subsection{Linearised CBE}

On dynamical timescales the DF of a  stellar system satisfies the
collisionless Boltzmann equation
\[
{\p f\over\p t}+[f,H]=0.
\]
When we split $f(\vtheta,\vJ,t)=f_0(\vJ)+f_1(\vtheta,\vJ,t)$ into its mean-field
and fluctuating components and neglect terms quadratic and higher in the
fluctuations, the CBE can be written
\[\label{eq:LCBE}
{\p f_1\over\p t}+[f_1,H_0]+[f_0,\Phi_1]=0.
\]
When we use angle-action coordinates to evaluate the Poisson brackets and
write $f_1$ and $\Phi_1$ as Fourier series in angles, equation  \eqrf{eq:LCBE}
yields 
\[\label{eq:nLCBE}
{\p f_\vn\over\p t}+\i\vn\cdot\vOmega f_\vn-\i\vn\cdot{\p
f_0\over\p\vJ}\Phi_{1\vn}=0,
\]
where we have dropped the subscript 1 from $f_{1\vn}$ for brevity but
retained it on $\Phi_{1\vn}$ for reasons that will soon become apparent.

\section{Energy of a perturbation}\label{sec:E}

Following \cite{NelsonTremaine1999} we imagine using an externally applied
gravitational field to impose a real perturbation $f_1$ on a mean-field model
$f_0(\vJ)$. The
perturbing gravitational potential $\Phi_1$ now has two components, the
potential $\Phi_\e$ of the externally applied field and the potential
$\Phi_\s$ generated via Poisson's equation by the perturbed density
distribution 
\[
\rho_1=\int\d^3\vv\,f_1.
\]
 
The rate at which work is done by the external field is 
\begin{align}\label{eq:E1}
{\d E\over\d t}&=-\int\d^3\vx\,\vv\cdot{\p\Phi_\e\over\p\vx}\rho_1\cr
&=-\int\d^6\vw\,f_1(\vw)\vv\cdot{\p\Phi_\e\over\p\vx}.
\end{align}
Now 
\[
\vv\cdot{\p\Phi_\e\over\p\vx}=-[H_0,\Phi_\e]=\vOmega\cdot{\p\Phi_\e\over\p\vtheta},
\]
so equation \eqrf{eq:E1} can be written
\begin{align}
{\d E\over\d t}&=-\int\d^6\vw\,f_1(\vw)\vOmega\cdot{\p\Phi_\e\over\p\vtheta}\cr
&=-(2\pi)^3\int\d^3\vJ\sum_\vn\,f^*_\vn(\vw)\i\vOmega\cdot\vn\,\Phi_{\e\vn}.
\end{align}
With $\Phi_1$ decomposed into its two components, the linearised CBE
\eqrf{eq:nLCBE} can be written
\[\label{eq:itsok}
{\p f_\vn\over\p t}+\i\vn\cdot\vOmega f_\vn-\i\vn\cdot{\p
f_0\over\p\vJ}\Phi_{\s\vn}=\i\vn\cdot{\p
f_0\over\p\vJ}\Phi_{\e\vn}.
\]
Eliminating $\Phi_{\e\vn}$ between the last two equations
\begin{align}\label{eq:dEdt}
{\d E\over\d t}\equiv-(2\pi)^3&\int\d^3\vJ\sum_\vn\,f^*_\vn{\vn\cdot\vOmega\over\vn\cdot\vnabla_\vJ
f_0}\,\cr
&\times\bigg({\p f_\vn\over\p t}+\i\vn\cdot\vOmega f_\vn-\i\vn\cdot\vnabla_\vJ
f_0\Phi_{\s\vn}.
\bigg)
\end{align}
The integrand contains three terms. The middle term is proportional to
$\vn\cdot\vOmega|f_\vn|^2$ and vanishes when summed over $\vn$ because the sum
includes both $\vn$ and $-\vn$.  The integral over the first term yields
\[
{\d K\over\d t}=-{(2\pi)^3\over2}\int\d^3\vJ\,\sum_\vn
{\vn\cdot\vOmega\over\vn\cdot\vnabla_\vJ f_0}
{\p|f_\vn|^2\over\p t}.
\]
In preparation for handling the third term, we note that
\begin{align}
\int\d^3\vtheta\,
f_1[\Phi_\s,H_0]&=\int\d^3\vtheta\,f_1{\p\Phi_\s\over\p\vtheta}\cdot\vOmega\cr
&=(2\pi)^3\sum_\vn f_{\vn}^*\i\vn\cdot\vOmega\Phi_{\s \vn}.
\end{align}
Hence the third term in equation \eqrf{eq:dEdt} is
\begin{align}
{\d P\over\d
t}&\equiv(2\pi)^3\int\d^3\vJ\sum_\vn\i\vn\cdot\vOmega f_\vn^*\Phi_{\s\vn}\cr
&=\int\d^6\vw\,f_1[\Phi_\s,H_0]=\int\d^6\vw\,f_1\vv\cdot{\p\Phi_\s\over\p\vx}\cr
&=\int\d^3\vx\,\Phi_\vs{\p\rho_1\over\p t}
=-{\d\over\d t}\fracj12G\int\d^3\vx\,\d^3\vx'\,{\rho_1(\vx)\rho_1(\vx')\over|\vx-\vx'|}
\end{align}
where the penultimate equality uses integration by parts and the continuity
equation $\p\rho/\p t=-\vnabla_\vx\cdot(\rho\vv)$.

Now that the two surviving contributions to the integral in equation
\eqrf{eq:dEdt} have proved to be total time derivatives, we can immediately
integrate from $E=K=P=0$ at $t=0$ to obtain an equation for the energy of an arbitrary
fluctuation
\begin{align}\label{eq:Exx}
E=-{(2\pi)^3\over2}\int\d^3\vJ\,\sum_\vn
{\vn\cdot\vOmega\over\vn\cdot\vnabla_\vJ f_0}
|f_\vn|^2
-\fracj12\int\d^3\vx\,\d^3\vx'\,{G\rho_1^2\over|\vx-\vx'|}.
\end{align}
This equation {\rd generalises to
three dimensions the quantity \cite{Kalnajs1971} shows to be
constant in an isolated razor-thin disc (Kalnajs' eqn.~48), and which he
says he will
`identify as energy'. Our derivation shows that $E$ is the work
that must be done to establish a perturbation rather than just showing that
$E$ is constant when $\Phi_\e$ vanishes.}  Equation \eqrf{eq:Exx}
differs from equation (5.130) in \cite{GDII} for the energy of a disturbed
ergodic system only by the replacement of $\d f_0/\d H$ by
$\vn\cdot\vnabla_\vJ f_0/\vn\cdot\vOmega$. When equation \eqrf{eq:Gxmx} is
used to eliminate $|\vx-\vx'|$ from equation \eqrf{eq:Exx} we obtain
\[\label{eq:EAA}
E=-{(2\pi)^3\over2}\int\d^3\vJ\,\sum_\vn
{\vn\cdot\vOmega\over\vn\cdot\vnabla_\vJ f_0}|f_\vn|^2
-\fracj12\cE\sum_\alpha|A_\alpha|^2.
\]

{\rd
\subsection{Restriction of the perturbed DF}\label{sec:restrict}

For certain combinations of $\vn$ and $\vJ$, $\vn\cdot\vnabla_\vJ f_0$ will
vanish and one might worry that such combinations will make the integral in
equation \eqrf{eq:Exx} for $E$ ill defined. However, from equation \eqrf{eq:itsok} it
follows that for these combinations $f_\vn(\vJ)$ remains zero as the
disturbance is excited, so the integrand in equation \eqrf{eq:Exx} vanishes
at the apparently problematic points. Thus the disturbances $f$ that can be
induced in a system with equilibrium DF $f_0(\vJ)$ by an external potential
are restricted in form. 

\cite{Antonov1961} broadened this result by showing \citep[e.g.][p.~429]{GDII} that the the perturbation
$f_1$ that is generated by applying {\it any} disturbing Hamiltonian $H_1$ to
$f_0$ will be of the form
\[
f_1=[h,f_0]={\p h\over\p\vtheta}\cdot{\p f_0\over\p\vJ},
\]
where $h(\vw)$ is a function that depends on $H_1$. On expanding $f$ and $h$
in Fourier series we obtain
\[
f_\vn=\i\vn\cdot{\p f_0\over\p\vJ}h_\vn
\]
so when $\vn\cdot\vnabla_\vJ f_0=0$, $f_\vn$ vanishes no matter how the system is
disturbed.
}

\subsection{A {\rd bilinear form} for DFs}

Additivity of energies is a fundamental property of normal modes. It emerges
naturally if we can express the energy of a disturbance as the norm of the
disturbance defined by a {\rd bilinear form} under which normal modes are mutually
orthogonal. Equation \eqrf{eq:EAA} suggests that the required form
is
\[\label{eq:defIP}
\bra{f}\widetilde f\ra=-{(2\pi)^3\over2}\int\d^3\vJ\,\sum_\vn
{\vn\cdot\vOmega\over\vn\cdot\vnabla_\vJ f_0}
f^*_\vn \widetilde f_\vn
-\fracj12\cE\sum_\alpha A^*_\alpha A'_\alpha,
\]
where $A_\alpha[f]$ is a functional of $f$ while $A'_\alpha$ denotes the
corresponding functional of $\widetilde f$.

\section{Normal modes of a non-ergodic system}\label{sec:mode}

We now look for disturbances with exponential time dependence, so
\[
f_\vn(\vJ,t)=f_\vn(\vJ,\omega)\e^{-\i\omega t},\quad
\Phi_\vn(\vJ,t)=\Phi_\vn(\vJ,\omega)\e^{-\i\omega t}
\]
With this ansatz, the linearised CBE \eqrf{eq:nLCBE} becomes
\[
(\vn\cdot\vOmega-\omega)f_\vn=(\vn\cdot\vnabla_\vJ f_0)\Phi_\vn(\vJ,\omega).
\]
This equation yields a well defined value for $f_\vn$ in terms of
$\Phi_\vn$ when $\vn\cdot\vOmega\ne\omega$, but on resonant tori (tori on
which $\vn\cdot\vOmega=0$) it
does not constrain $f_\vn$. Therefore, as \cite{vanKampen1955} pointed out,
the solution to this equation must be written
\[\label{eq:DFvK}
f_\vn(\vJ,\omega)=\cP{\vn\cdot\vnabla_\vJ f_0\over\vn\cdot\vOmega-\omega}
\Phi_\vn(\vJ,\omega)+g_\vn(\vJ)\delta(\vn\cdot\vOmega-\omega),
\]
 where $\cP$ indicates that when integrated wrt $\vn\cdot\vOmega$ or
 $\omega$, the Cauchy principal
value around the singularity at $\omega=\vn\cdot\vOmega$ should be taken, and
$g_\vn(\vJ)$ is an arbitrary function that specifies non-trivial
distributions of stars on the resonant tori. {\rd In Section
\ref{sec:restrict} we saw that the DFs of disturbances are restricted such
than $f_\vn=0$ when $\vn \cdot \vnabla_\vJ f_0=0$. Since a system's normal
modes comprise possible  disturbances, $g_\vn$ must be restricted in the
same way:
\[
\vn\cdot\vnabla_\vJ f_0=0\quad\Rightarrow\quad g_\vn=0.
\]
}

In Paper I
$\Phi_\vn(\vJ,\omega)$ in equation \eqrf{eq:DFvK} is interpreted as the
potential generated by the the resonant stars and by the non-resonant stars
that they excite. It is the {\it dressed} potential of the resonant stars.
Consequently, Poisson's equation imposes a relation between $g_\vn$ and
$\Phi_\vn$. We obtain this relation by multiplying equation \eqrf{eq:DFvK} by
$\d^6\vw\,\e^{\i\vn\cdot\vtheta}\Phi^{(\alpha')*}$ and integrating through
phase space to obtain the self-consistency condition (cf Paper I eqn.~42)
\[\label{eq:MAeqB}
\sum_\alpha M_{\alpha'\alpha}A_{\alpha}=-B_{\alpha'}
\]
where
\begin{align}
M_{\alpha'\alpha}(\omega)&=\delta_{\alpha'\alpha}-{(2\pi)^3\over\cE}\cP\int\d^3\vJ\,
\sum_\vn{\vn\cdot\vnabla f_0\over\vn\cdot\vOmega-\omega}\Phi_\vn^{(\alpha')*}\Phi_\vn^{(\alpha)}\cr
B_{\alpha'}&=-{(2\pi)^3\over\cE}\int\d^3\vJ\,\sum_\vn
g_\vn(\vJ)\Phi_\vn^{(\alpha')*}(\vJ)\delta(\vn\cdot\vOmega-\omega).
\end{align}
In any truly stable system \citep[so excluding marginally stable systems such
as those discussed by][]{Mathur1990}, $M(\omega)$ has an inverse for all real
$\omega$, so given $g(\vw)$ we can solve for the amplitudes $A_\alpha$ of the
corresponding density and potential. The coefficients $B_{\alpha'}$ give the
density and potential generated by the resonant stars alone, that is, after
removing the contributions of stars driven by the gravitational field of the
resonant stars (Paper I eqn.~46).

In principle for given functions $g_\vn$ equation \eqrf{eq:DFvK} yields a
mode for every frequency of the form $\vn\cdot\vOmega$, but since vectors
$\vn$ with any large component generate very small densities in real space,
the important modes are confined to frequencies that range from near zero
($|\vOmega|\approx0$ at large $|\vJ|$) to about twice the system's maximum
circular frequency.

$B$ necessarily vanishes for complex $\omega$ because  $\vn\cdot\vOmega$ is
inherently real. Thus all van Kampen modes have real frequencies.

Any additional modes must arise when $|\vM(\omega)|=0$ at some possibly
complex frequency $\omega$.  If the underlying equilibrium $f_0$ is unchanged
by reversing all velocities (in practice meaning that it has no net
rotation), then if $|\vM(\omega)|=0$ then $|\vM(-\omega)|=0$ also, and at one
of these two frequencies the disturbance will grow exponentially and the
system is unstable.  Hence every mode of a truly stable, time-reversible
system is a van Kampen mode. This result extends to flattened systems and
systems with velocity anisotropy a key result of Paper I.
We defer discussion of marginally stable systems to Section
\ref{sec:discuss}.

The $\cP$ and $\delta$ symbols in equation \eqrf{eq:DFvK} signal that van
Kampen modes  make sense only within an integral wrt $\omega$. Hence a
physical disturbance will always be of the form
\[\label{eq:decomp}
F(\vtheta,\vJ,t)=\sum_\vn\int\d\omega\,f_\vn(\vJ,\omega)\e^{\i(\vn\cdot\vtheta-\omega
t)}.
\]
This disturbance is specified by one's choice of the functions $g_\vn(\vJ)$.

\subsection{Energies of normal modes}

In Appendix A we show that in the case of two normal modes $f(\omega)$ and
$\widetilde f(\widetilde\omega)$ the {\rd bilinear form} \eqrf{eq:defIP} can be brought to
the form
\begin{align}\label{eq:vKIP}
\bra{f}\widetilde f\ra=-{(2\pi)^3\over2}&\sum_\vn
\int\d^3\vJ\,\omega
\bigg\{\pi^2(\vn\cdot\vnabla_\vJ f_0)
\Phi_\vn^*\widetilde\Phi_\vn\cr
&+{g^*_\vn\widetilde g_\vn\over\vn\cdot\vnabla_\vJ f_0}\bigg\}
\delta(\vn\cdot\vOmega-\omega)\delta(\omega-\widetilde\omega).
\end{align}
Remarkably, on account of the factor $\delta(\vn\cdot\vOmega-\omega)$ the
{\rd bilinear form} is computed by integrating only over resonant stars, despite
much of the energy lying with driven, non-resonant stars.  Combining this
equation for the {\rd bilinear form evaluated on} two van Kampen modes with the
equation \eqrf{eq:decomp} expressing a general disturbance as a sum of van
Kampen modes, we obtain an alternative expression for the value of the form
on any two
disturbances $F$ and $\widetilde F$:
\begin{align}\label{eq:otherIP}
\bra{F}&\widetilde F\ra=\Big<{\int\!\d\omega\,
f(\omega)}\Big|\int\!\d\widetilde\omega\,\widetilde f(\widetilde\omega)\Big\ra
=\int\!\d\omega\,\d\widetilde\omega\,\bra{f(\omega)}\widetilde f(\widetilde\omega)\ra\cr
&=-{(2\pi)^3\over2}\sum_\vn
\int\d^3\vJ\,\omega_\vn(\vJ)\cr
&\times\bigg\{\pi^2(\vn\cdot\vnabla_\vJ
f_0)\Phi_\vn^*(\omega_\vn)\widetilde\Phi_\vn(\omega_\vn)
+{g_\vn^*(\omega_\vn)\widetilde g_\vn(\omega_\vn)\over\vn\cdot\vnabla_\vJ f_0}\bigg\},
\end{align}
where $\omega_\vn(\vJ)\equiv\vn\cdot\vOmega$ and the potentials $\Phi_\vn$ and
driving terms $g_\vn$ have acquired $\omega_\vn$ as an argument to indicate that
they are the potentials and driving terms of the van Kampen modes with that
frequency in the decomposition of $F$ into modes.

When we set $\widetilde F=F$ in equation \eqrf{eq:otherIP} we obtain  the energy of a general
disturbance as the sum of the energies of its component van Kampen modes:
\[\label{eq:EF}
E[F]=\bra FF\ra=(2\pi)^3\sum_\vn\int\d^3\vJ\,\omega_\vn(\vJ)N_\vn(\vJ),
\]
where
\[
N_\vn(\vJ)=-\fracj12\Big\{\pi^2(\vn\cdot\vnabla_\vJ f_0)
 |\Phi_\vn(\omega_\vn)|^2
+{|g_\vn(\omega_\vn)|^2\over\vn\cdot\vnabla_\vJ f_0}\Big\}
\]
is the phase-space density of `plasmons' associated with wavevector $\vn$ in the terminology of
\cite{HamiltonHeinemann2020}.
Remarkably, this expression for the plasmon density involves only the driving
term $g_\vn$ and the resulting spatial structure $\Phi_\vn$ of the component
van Kampen modes -- there is no mention of the kinetic energy of the driven stars.
At first sight this is odd because in a
stable system disturbances fade while $E$ is constant. The equation works
because the potentials in question belong not to the disturbance but to its
constituent van Kampen modes, which do not change but nevertheless cause the
disturbance to fade as they become more and more evenly distributed in phase.
 
\section{Discussion}\label{sec:discuss}

There are close parallels between formulae derived here for general systems
and ones presented in Paper I for ergodic systems. The essential difference
is the universal replacement of $\d f_0/\d H$ in Paper I by
$\vn\cdot\vnabla_\vJ
f_0/\vn\cdot\vOmega$ as here. The other differences are more superficial, being
caused by formulae in Paper I being derived from an operator that gives $\p^2
f/\p t^2$ rather than one that gives $\p f/\p t$. This change leads to
$\delta\big((\vn\cdot\vOmega)^2-\omega^2\big)$ in Paper I being replaced by
$\delta(\vn\cdot\vOmega-\omega)$. These symbols have different dimensions:
\[
\delta\big((\vn\cdot\vOmega)^2-\omega^2\big)
=\delta\big((\vn\cdot\vOmega-\omega)(\vn\cdot\vOmega+\omega)\big)
\simeq{\delta(\vn\cdot\vOmega-\omega)\over2\omega}.
\]
As a consequence, $g_\vn/2\omega$ in Paper I is equivalent to $g_\vn$ here.

In Paper I we explored the consequences of reducing the self-gravity of a
system by reducing the masses of its particles by a factor $\xi<1$ and
introducing a fraction $(1-\xi)$ of the mean-field potential. This operation
modified the structure of the system's van Kampen modes by suppressing the
dressing of the potential of resonant stars. The ensuing discussion  applies equally to
the general equilibria treated here.

In Paper I we stressed that Landau `modes' are not really modes but zeroes of
the function $|\vM(\omega)|$. When such a zero lies just below the real axis
(a `weakly damped mode') the van Kampen modes on the adjacent stretch of the
real axis are heavily dressed and will make a large contribution to the
system's evolution. The time required for these modes to drift out of phase,
and the associated disturbance to fade in real space, is inversely
proportional to extent of the heavily dressed section of the real line, and
therefore inversely proportional to the distance of the zero from the real
line. This interpretation of `Landau damping' applies equally to the general
systems discussed here.

{\rd Although \cite{Mathur1990} did not display an explicit example, he
demonstrated the logical possibility of marginally stable systems, that is
ones that can oscillate for ever. This possibility arises when frequencies of
the form $\vn\cdot\vOmega(\vJ)$ do not extend from a maximum frequency down
to zero. The easiest way to engineer a gap in the frequency coverage is to
impose a limit on the spatial extent of the system, so there is a minimum
orbital frequency $\Omega_{\rm min}$, and to make the system effectively
one-dimensional so small frequencies cannot be constructed by differencing
frequencies greater than $\Omega_{\rm min}$. When these conditions are
satisfied, \cite{Mathur1990} shows that $f_0$ can be devised such that the
matrix $\vM$ of equation \eqrf{eq:MAeqB} has vanishing determinant in a gap.
Consequently, non-zero $\vA$ can then be found even though $\vB$ vanishes because we
are in a gap. In these exceptional circumstances, a complete set of modes
comprises the van Kampen modes plus any stable/unstable pairs of modes and
any marginally stable modes in gaps.}

\section{Conclusions}\label{sec:conclude}

We have extended results presented in Paper I from ergodic systems to
systems with DFs of the form $f(\vJ)$. This is a major extension because
ergodic systems probably do not occur in Nature while many galaxies and star
clusters will have DFs that can be closely approximated by $f(\vJ)$.

We derived equation \eqrf{eq:Exx} for the energy of a disturbance to a
general system, and motivated by this result defined a {\rd bilinear form} on the
space of DFs such that the energy of a disturbance is the norm of its DF.

Next we extended the concept of van Kampen modes to general systems.  They
exist for essentially the same range of real frequencies, and have the same
physical interpretation and energy-additivity as in the ergodic case. There
is again a sharp distinction between van Kampen modes, which have frequencies
that lie in a real continuum, and classical modes, which have isolated, and
{\rd generally
complex,} frequencies. When a system is time-reversible (lacks rotation) and
stable, it can only have van Kampen modes.

The mathematical apparatus used here is simpler and less powerful than that
deployed in Paper I. On the plus side, the formulae presented here are
somewhat simpler than the corresponding formulae in Paper I. The downsides
are (i) that modes do not
emerge as eigenfunctions of a Hermitian operator; (ii) we have not shown that all van Kampen modes have positive
energy. 
The lack of a connection to a Hermitian operator deprives us of the ability
both to argue for the completeness of the modes and 
to confine the normal-mode frequencies to the
axes of the complex plane as we could in the case of ergodic systems.
Nonetheless, at least in the case of systems with $f_0(\vJ)$ that are
unchanged by reversing all velocities, it seems likely that the frequencies
are confined in the same way. This topic will be a major theme of Paper IV in
this series.

The existence of modes with negative energy would not have important
consequences on the dynamical timescale, because modes evolve independently
of one another so long as the linear approximation holds. On the longer
timescales associated with terms quadratic in the disturbance (the `two-body'
timescale), a system {\rd is likely to} be rendered secularly unstable by the existence of
negative-energy modes because non-linear terms could transfer energy from
negative-energy modes to positive-energy terms and thus causing the
amplitudes of both types of mode to increase.

\section*{Acknowledgements}

We thank Scott Tremaine for an insightful referee's report.
Jun Yan Lau
gratefully acknowledges support from University College London's Overseas and
Graduate Research Scholarships.  James Binney is supported by the UK Science
and Technology Facilities Council under grant number ST/N000919/1 and by the
Leverhulme Trust through an Emeritus Fellowship. 

\section*{Data Availability}

No new data was generated or analysed in support of this research.

\bibliographystyle{mn2e} 
\bibliography{new_refs.bib}

\appendix
\section{The {\rd bilinear form evaluated on} van Kampen modes}\label{sec:A}

Here we prove that the {\rd bilinear form \eqrf{eq:defIP} evaluated on} two van
Kampen modes can be expressed in the form \eqrf{eq:vKIP}. The product is the
sum $K+\frac12P'$ of kinetic- and potential-energy terms, where\footnote{$P'$
is the potential energy of the system defined by $f$ in the potential
generated by $\widetilde f$ or vice versa [eqn.~\ref{eq:PE3}]. The
self-energy $P=\fracj12P'[f,f]$.}
\begin{align}
K[f,\widetilde f]&\equiv-{(2\pi)^3\over2}\sum_\vn\int\d^3\vJ\,
{\vn\cdot\vOmega\over\vn\cdot\vnabla_\vJ f_0}
f^*_\vn\widetilde f_\vn\cr
P'[f,\widetilde f]&\equiv-\cE\sum_\alpha A^*_\alpha\widetilde A_\alpha.
\end{align}
Using equation \eqrf{eq:DFvK} for the DF of a mode, we have
\begin{align}\label{eq:defK}
K=-&{(2\pi)^3\over2}\sum_\vn\int\d^3\vJ{\vn\cdot\vOmega\over\vn\cdot\vnabla_\vJ
f_0}\cr
&\ \times\bigg\{\cP\Big({\vn\cdot\vnabla_\vJ
f_0\over\vn\cdot\vOmega-\omega}\Big)\Phi^*_\vn+g^*_\vn\delta(\vn\cdot\vOmega-\omega)
\bigg\}\cr
&\ \times\bigg\{\cP\Big({\vn\cdot\vnabla_\vJ
f_0\over\vn\cdot\vOmega-\widetilde\omega}\Big)\widetilde\Phi_\vn+\widetilde
g_\vn\delta(\vn\cdot\vOmega-\widetilde\omega).
\bigg\}
\end{align}
With identities \eqrf{eq:ID1} and \eqrf{eq:ID2} the product of the two
principal values yields
\begin{align}\label{eq:PP}
\int\d^3\vJ\,&{\vn\cdot\vOmega\over\vn\cdot\vnabla_\vJ f_0}\cP(.)\cP(.)=
\int\d^3\vJ\,\vn\cdot\vnabla_\vJ f_0\Phi^*_\vn\widetilde\Phi_\vn\cr
&\times\bigg[\cP\Big({1\over\omega-\widetilde\omega}\Big)\bigg\{
\cP\Big({\omega\over\vn\cdot\vOmega-\omega}\Big)
-\cP\Big({\widetilde\omega\over\vn\cdot\vOmega-\widetilde\omega}\Big)\bigg\}\cr
&+\pi^2\omega\delta(\vn\cdot\vOmega-\omega)\delta(\omega-\widetilde\omega)\bigg].
\end{align}
The cross terms yield
\begin{align}
\int\d^3\vJ\,\Big\{&
\cP\Big({\vn\cdot\vOmega\over\vn\cdot\vOmega-\omega}\Big)\Phi^*_\vn \widetilde
g_\vn\delta(\vn\cdot\vOmega-\widetilde\omega)\cr
&+\cP\Big({\vn\cdot\vOmega\over\vn\cdot\vOmega-\widetilde\omega}\Big)\widetilde\Phi_\vn 
g^*_\vn\delta(\vn\cdot\vOmega-\omega).
\Big\}
\end{align}
The identity \eqrf{eq:ID3}
enables us to rewrite this in the form
\begin{align}\label{eq:Pg}
\int\d^3\vJ\,\cP\Big({1\over\omega-\widetilde\omega}\Big)&\int\d^3\vJ\,\Big\{
\widetilde\omega\Phi^*_\vn \widetilde
g_\vn\delta(\vn\cdot\vOmega-\widetilde\omega)\cr
&-\omega\widetilde\Phi_\vn 
g^*_\vn\delta(\vn\cdot\vOmega-\omega)
\Big\}.
\end{align}
When  the cross terms are added to the reduced product of principal values
\eqrf{eq:PP}, the coefficient of $\cP\big(1/(\omega-\widetilde\omega)\big)$ is
\begin{align}
\Big\{\cP\Big({\vn\cdot\vnabla_\vJ
f_0\over\vn\cdot\vOmega-\omega}&\Big)\Phi_\vn^*+g^*_n\delta(\vn\cdot\vOmega-\omega)\Big\}\omega\widetilde\Phi_\vn\cr
-&\Big\{\cP\Big({\vn\cdot\vnabla_\vJ
f_0\over\vn\cdot\vOmega-\widetilde\omega}\Big)\widetilde\Phi_\vn+\widetilde
g_n\delta(\vn\cdot\vOmega-\widetilde\omega)\Big\}\widetilde\omega\Phi^*_\vn\cr
&\hskip1cm=f_\vn^*\omega\widetilde\Phi_\vn-\widetilde f_\vn\widetilde\omega\Phi^*_\vn.
\end{align}
When this expression is integrated wrt $\vJ$, summed over $\vn$, and
multiplied by $(2\pi)^3$, we obtain $\omega-\omega'$ times the mutual
potential energy $P'$ (eqn.~\ref{eq:PE3}). Hence the denominator of the
principal value symbol in equation \eqrf{eq:Pg} is cancelled. Gathering
together the
contributions to $K$, we have this potential-energy term, the 
product of $\delta$-functions in equation \eqrf{eq:PP} and a similar term
arising from the product of $g^*_\vn$ and $\widetilde g_\vn$ in equation
\eqrf{eq:defK}. Thus
\begin{align}
K=-\fracj12P'&-{(2\pi)^3\over2}\sum_\vn\int\d^3\vJ\,
\omega\Big(\pi^2\vn\cdot\vnabla_\vJ f_0\Phi^*_\vn\widetilde\Phi_\vn\cr
&+{\vn\cdot\vOmega\over\vn\cdot\vnabla_\vJ f_0}g^*_\vn\widetilde g_\vn\Big)
\delta(\vn\cdot\vOmega-\omega) \delta(\omega-\widetilde\omega).
\end{align}
Thus $\bra f\widetilde f\ra=K+\fracj12P'$ is indeed given equation \eqrf{eq:vKIP}.

\section{Identities involving principal values}\

Here we list three identities that are required in Appendix \ref{sec:A}. For proofs
see  \citep[e.g.][]{RamosWhite2018}.
\begin{align}\label{eq:ID1}
\cP\Big({1\over x-x_1}\Big)&\cP\Big({1\over x-x_2}\Big)=\cr
&\cP\Big({1\over x_1-x_2}\Big)\bigg\{
\cP\Big({1\over x-x_1}\Big)-\cP\Big({1\over x-x_2}\Big)\bigg\}\cr
&+\pi^2\delta(x-x_1)\delta(x_1-x_2),
\end{align}
\[\label{eq:ID2}
\cP\Big({x\over x-x_1}\Big)=1+\cP\Big({x_1\over x-x_1}\Big).
\]
\begin{align}\label{eq:ID3}
 \cP\Big(&{A\over x - x_2}\Big)\delta(x - x_1) + \cP\Big({B\over x -
 x_1}\Big)\delta(x - x_2) \cr
&= \cP\Big({1\over x_1 - x_2}\Big)\big\{A\delta(x - x_1) - B\delta(x -
x_2)\big\}.
\end{align}

\end{document}